\documentclass[useAMS,usenatbib]{mn2e}

\usepackage{natbib}
\usepackage{graphicx}
\usepackage{amssymb}

\title[Radio-mode feedback in local AGNs]{Radio-mode feedback in local AGNs: dependence on the central black hole parameters}
\author[W. Ishibashi et al.]
{W. Ishibashi$^{1,}$$^{2}$\thanks{E-mail:
wako.ishibashi@phys.ethz.ch}, M. W. Auger$^{1}$, D. Zhang$^{1}$, and A.C. Fabian$^{1}$  \\
\footnotemark[0]\\
$^{1}$Institute of Astronomy, Madingley Road, Cambridge CB3 0HA 
\footnotemark[0]\\
$^{2}$Institute for Astronomy, Department of Physics, ETH Zurich, Wolfgang-Pauli-Strasse 27, CH-8093 Zurich, Switzerland } 

\begin{document}

\pdfminorversion=4

\date{Accepted ? Received ?; in original form ? } 

\pagerange{\pageref{firstpage}--\pageref{lastpage}} \pubyear{2012}

\maketitle

\label{firstpage}

\begin{abstract}
Radio mode feedback, in which most of the energy of an active galactic nucleus (AGN) is released in a kinetic form via radio-emitting jets, is thought to play an important role in the maintenance of massive galaxies in the present-day Universe. We study the link between radio emission and the properties of the central black hole in a large sample of local radio galaxies drawn from the Sloan Digital Sky Survey (SDSS), based on the catalogue of \citet{B_H_2012}.
Our sample is mainly dominated by massive black holes (mostly in the range $10^8-10^9 M_{\odot}$) accreting at very low Eddington ratios (typically $\lambda < 0.01$). In broad agreement with previously reported trends, we find that radio galaxies are preferentially associated with the more massive black holes, and that the radio loudness parameter seems to increase with decreasing Eddington ratio. We compare our results with previous studies in the literature, noting potential biases. The majority of the local radio galaxies in our sample are currently in a radiatively inefficient accretion regime, where kinetic feedback dominates over radiative feedback. We discuss possible physical interpretations of the observed trends in the context of a two-stage feedback process involving a transition in the underlying accretion modes. 
\end{abstract} 



\section{Introduction}

Accreting black holes at the centre of galaxies release huge amounts of energy into the surroundings in both radiative and kinetic forms. 
The interaction between this `feedback' and the ambient material can have a strong impact on the host galaxy, and AGN (active galactic nucleus) feedback is now thought to play a major role in galaxy evolution. 
Observations show a number of important correlations between the central black hole and its host galaxy, such as the black hole mass-velocity dispersion relation (the so-called $M - \sigma$ relation) and the black hole mass-bulge stellar mass relation \citep{Magorrian_et_1998, Gebhardt_et_2000, Ferrarese_Merritt_2000, Marconi_Hunt_2003, Haring_Rix_2004}. 
The existence of such tight correlations indicates a close coupling between the central black hole and its host galaxy, suggesting some form of `co-evolution'. 
The observed correlations are generally interpreted in the framework of AGN feedback models \citep[][and references therein]{Silk_Rees_1998, Fabian_1999, King_2003, Murray_et_2005, Fabian_2012}. 
Two main modes of feedback are observed in nature: the radiative (or quasar) mode and the kinetic (or radio) mode, which possibly operate at different stages in the evolution of galaxies \citep[][]{Fabian_2012}. 

AGN feedback is usually invoked to suppress star formation in the host galaxy, either by removing or heating the ambient gas in the radiative and kinetic modes, respectively. 
The quasar mode is associated with radiatively efficient accretion approaching the Eddington limit, and is most active at high redshifts ($z\sim 2$), close to the peak epoch of AGN activity. 
In contrast, in the radio mode, much of the energy output is in kinetic form and channeled through relativistic radio jets. 
The kinetic mode is generally associated with lower accretion rates and mainly operates at lower redshifts. 
The mechanical energy deposited by the jet can heat the ambient gas and prevent radiative cooling, thus inhibiting star formation, and limiting the stellar mass of the host galaxy. 
The required balance between heating and cooling rates seems to be preserved in this maintenance mode, which shapes the late evolution of massive galaxies. 
In the present-day Universe, radio mode feedback typically operates in massive galaxies with low accretion rates, and is commonly observed in local giant ellipticals. 

The relative importance of radio emission, compared to the other wavebands, can vary significantly in different sources. Accordingly, AGNs can be divided into two distinct classes: radio-loud (RL) and radio-quiet (RQ), leading to the well-known radio-loud/radio-quiet dichotomy. 
The distinction between the two classes is based on the radio loudness of the source, usually quantified by the radio loudness parameter $R = \frac{F_r}{F_o}$, defined as the ratio of monochromatic luminosities at 5 GHz and optical B-band at 4400$\AA$ \citep{Kellermann_et_1989}. 
The formal boundary between the radio-loud and radio-quiet objects is set at $R \sim 10$ \citep{Kellermann_et_1994}. 
A bimodal distribution, formed by two distinct sequences, can be seen in the radio luminosity versus optical luminosity plane \citep[e.g.][]{Xu_et_1999, Sikora_et_2007}, although the existence of a real physical dichotomy in the radio loudness distribution of AGNs has been questioned \citep[][and references therein]{Falcke_et_1996, Lacy_et_2001, Balokovic_et_2012}. 

The relation between radio emission and the properties of the central accreting black hole, i.e. black hole mass and accretion rate, has been extensively studied in the literature for a variety of AGN samples selected following different selection criteria \citep{Laor_2000, Ho_2002, McLure_Jarvis_2004, Sikora_et_2007, Chiaberge_Marconi_2011, Sikora_et_2013, Wu_et_2013}.  
Two interesting trends seem to emerge by analysing the dependence of the radio loudness on the central black hole parameters: an anti-correlation with the Eddington ratio ($\lambda = L_{\rm bol}/L_{\rm Edd}$), and a possible correlation with the black hole mass ($M_{BH}$). In particular, \citet{Sikora_et_2007} report a double sequence structure in the radio luminosity vs. optical luminosity plane, in which both radio-loud and radio-quiet sequences are observed to follow the same trend of increasing radio loudness with decreasing Eddington ratio. 
Here we re-visit the issue based on a large sample of local radio-AGN drawn from the Sloan Digital Sky Survey (SDSS), and examine the dependence of the radio emission on the properties of the central black hole, with the aim of better characterising the occurrence of radio-mode feedback in the local Universe. 

The paper is organized as follows. In Sect. \ref{Section_Data} we introduce our sample of radio sources drawn from the SDSS, and describe the data analysis method in which we try to carefully take into account measurement uncertainties and error propagation. 
In Sect. \ref{Section_Results} we present the dependence of the radio emission on black hole mass and Eddington ratio, and in Sect. \ref{Section_Comparison} we compare our results with previous studies in the literature. We discuss a number of possible physical interpretations in Sect. \ref{Section_Discussion}.  


\section{Sample and data analysis}
\label{Section_Data}

We have compiled a sample of radio-loud AGNs derived from the catalogue provided by \citet{B_H_2012}, which was constructed by cross-matching radio sources from the National Radio Astronomy Observatory (NRAO) Very Large Array (VLA) Sky Survey 
\citep[NVSS;][]{Condon_et_1998} and the Faint Images of the Radio Sky at Twenty centimetres \citep[FIRST;][]{Becker_et_1995} survey with objects in the 7th data release (DR 7) of the Sloan Digital Sky Survey \citep[SDSS;][]{Abazajian_et_2009}. 
The original \citet{B_H_2012} catalogue includes 18286 sources. 
We use stellar velocity dispersion measurements and [Oiii] luminosities to derive black hole properties, and we therefore also only use objects with $\sigma <$ 400 km$\mathrm{s^{-1}}$ and require that the uncertainties on $\sigma$ and $L_{\rm Oiii}$ must be less than 10\% and 20\%, respectively; our final sample has 1091 objects.

We use the SDSS database and the FIRST/NVSS surveys to derive three measurements for each AGN: the [Oiii] luminosity, the stellar velocity dispersion, and the radio flux at 1.4 GHz, all with related uncertainties. 
We then apply a number of transformations to the data in order to derive quantities related to the central black hole. 

\subsection{Radio Luminosity}

The radio luminosity is computed using the standard luminosity-flux relation, with the radio flux measured at 1.4 GHz. 
The radio loudness parameter used in the literature is based on the radio luminosity at 5 GHz and the B-band luminosity at 4400 $\AA$, defined as $R \sim 10^5 (L_{\mathrm{5 GHz}}/L_B)$ \citep{Sikora_et_2007}, where the nuclear B-band luminosity is assumed to be a tracer of the accretion luminosity (see Sect. \ref{subsection_optical_luminosity}).  
The conversion from 1.4 GHz to 5 GHz luminosity is computed by assuming a radio spectral index of $0.9\pm0.35$ obtained from a Gaussian fit to the data provided by \citet{Randall_et_2012}.

\subsection{Optical Luminosity}
\label{subsection_optical_luminosity}

We do not have a direct measure of the B-band luminosity due to the presence of the AGN host galaxy, so in order to allow for comparisons with previous studies we also need to derive a relationship between the [Oiii] luminosity and the B-band luminosity. 
\citet{Shen_et_2011} report a correlation between the [Oiii] line luminosity and the continuum luminosity at 5100 $\AA$ for a sample of 105783 quasars in the SDSS (DR 7) quasar catalog. The best-fit relation is given by $\log (L(5100/\mathrm{erg s^{-1}}) = (0.771 \pm 0.002) \log(L(\mathrm{[Oiii])erg s^{-1}}) + (12.13 \pm 0.08)$. The 5100 $\AA$ luminosity is then converted into the 4400 $\AA$ B-band luminosity by assuming an optical spectral index of $\alpha_o = 0.5$. 
However, this relationship includes a large amount of scatter and also may introduce systematics since it is defined for quasars while our sample does not include any bright quasars and the continuum can be dominated by stellar light. 
Moreover, the power law index of $\alpha_o \sim 0.5$ at optical-UV wavelengths is a typical value obtained for luminous AGNs \citep[e.g][]{vandenBerk_et_2001}, while there are indications that the spectral slope might be steeper in lower luminosity sources \citep[][and references therein]{Ho_2008}. 
In fact, a characteristic feature of the spectral energy distribution of low-luminosity AGNs is the lack of the big blue bump component, with a steep optical-UV slope, although dust reddening effects have been invoked \citep{Maoz_2007}. 

On the other hand, the [Oiii] line luminosity is often used as a direct tracer of the AGN bolometric luminosity \citep[e.g.][]{Heckman_et_2004}. The [Oiii] emission line is produced by photoionisation of gas in the narrow line region, and is thus less affected by dust extinction due to the obscuring torus. But the optical line is also subject to extinction due to dust present in the interstellar medium of the host galaxy, and corrections for this effect should be taken into account \citep{Kauffmann_Heckman_2009}. 
In addition, while the [Oiii] luminosity is widely used as a proxy of the bolometric luminosity of radiative-mode AGN, it is not clear that it is a good indicator in the case of lower luminosity objects dominated by the kinetic output.  
One should therefore keep in mind these caveats when interpreting the results. 

\subsection{Black hole mass}

We derive the black hole mass from the stellar velocity dispersion by using the $M - \sigma$ relation determined by \citet{McConnell_Ma_2013}
$$
\rm{log} M_{\rm BH} = (8.32\pm0.05) + (5.58\pm0.34){\rm log}(\sigma) + 
(0.42\pm0.04)
$$
where $\sigma$ is in units of 200 km$\mathrm{s^{-1}}$. 
The Eddington ratio is given by
$$
\lambda = L_{\rm bol}/L_{\rm Edd}
$$
where $L_{\rm bol} = 10L_{\rm B}$ \citep{Richards_et_2006}. 
In Sect. \ref{Section_Results} we also estimate the bolometric luminosity directly from the \rm [Oiii] emission line, $L_{\rm bol} = 3500L_{\rm [Oiii]}$ \citep{Heckman_et_2004}.  

\subsection{Fitting Procedure}  

In practice, the propagation of the measurement uncertainties when performing these transformations can be non-trivial, in part because the transformations themselves can have uncertainties, e.g. in the $M - \sigma$ relation. We therefore use a Monte Carlo method to robustly propagate all of the known uncertainties by drawing 10000 samples from the measured distribution for $\sigma$, $L_{\rm 
[Oiii]}$, and $F_{\rm 1.4~GHz}$ for each AGN. We then transform these samples into $M_{\rm BH}$, $\lambda$, and $R$ using the above relations, including drawing samples for uncertain parameters as in the $M - \sigma$ relation, and find the mean and standard deviation of the transformed samples; we then take these to be our `measurements' for each 
of the derived quantities. Note that this method also allows us to determine the covariance between all of the quantities, both measured and derived. 
  
The purpose of this work is to look at the relationship between the radio luminosity and black hole properties of a large sample of radio-loud AGNs. We quantify these relationships by performing linear fits of one parameter against 
another, following the formalism of \citet{Kelly_2007}. In particular, this allows us to robustly include the uncertainties and correlations within the dataset, and we are also able to quantify the intrinsic scatter in the linear fits (i.e., an estimate of the true physical scatter deconvolved from the significant scatter induced by measurement uncertainties). It is important to note that attempting to determine the selection function for our dataset is intractable, but understanding selection biases is critical for interpreting the linear relationships that we are fitting. This problem is slightly alleviated by employing the \citet{Kelly_2007} scheme, which empirically determines the actual distributions from which our samples are drawn, although this is the product of the true distribution and the selection function. 

In the following, we split our radio-AGN sample into two sub-classes, high-excitation and low-excitation radio galaxies (HERG and LERG), according to the classification introduced by \citet{B_H_2012}. The resulting sample is composed of 890 LERGs and 179 HERGs. 


\section{Results}
\label{Section_Results}

\subsection{Dependence on the black hole mass}

\begin{figure}
\begin{center}
\includegraphics[angle=0,width=0.5\textwidth]{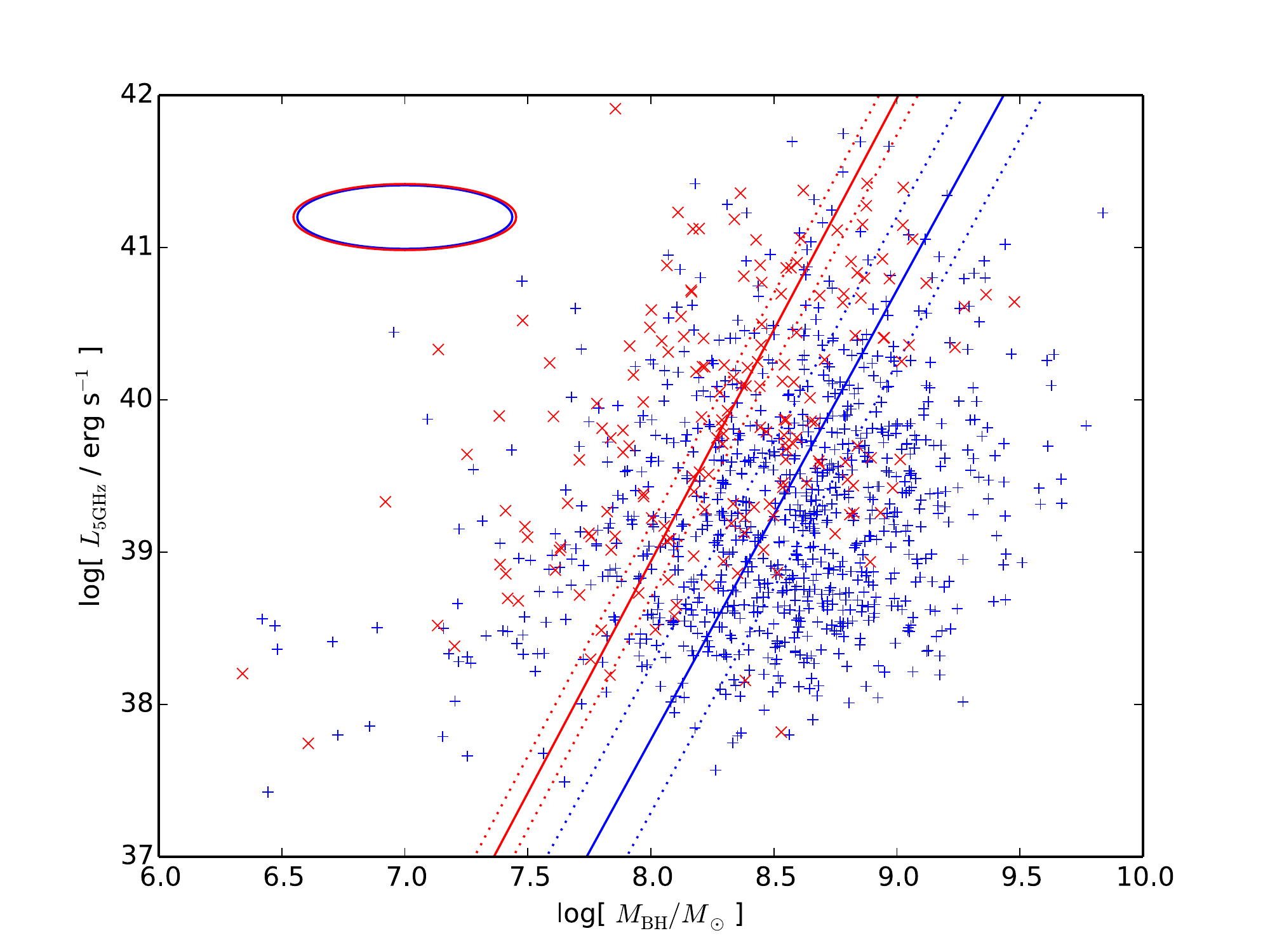} 
\caption{\small
The relationship between the 5 GHz radio luminosity and the black hole mass, with red crosses indicating HERGs and blue plus symbols indicating LERGs. 
The red solid line is our best-fit linear relation for HERGs, log $L_{R}$ = ($38.9\pm0.2$) + ($3.0\pm0.6$) [log $M_{BH} + 8]$, and the red dotted lines indicate the intrinsic scatter, which we find to be $0.24\pm0.15$. 
The blue solid line is the best-fit linear relation for LERGs, log $L_{R}$ = ($37.8\pm0.5$) + ($2.9\pm1.0$) [log $M_{BH} + 8]$, with the blue dotted lines indicating an intrinsic scatter of $0.48\pm0.12$. 
The ellipses in the upper left corner illustrate the typical uncertainty and covariance for each AGN.}
\label{Fig_LR_MBH}
\end{center}
\end{figure}

\begin{figure}
\begin{center}
\includegraphics[angle=0,width=0.5\textwidth]{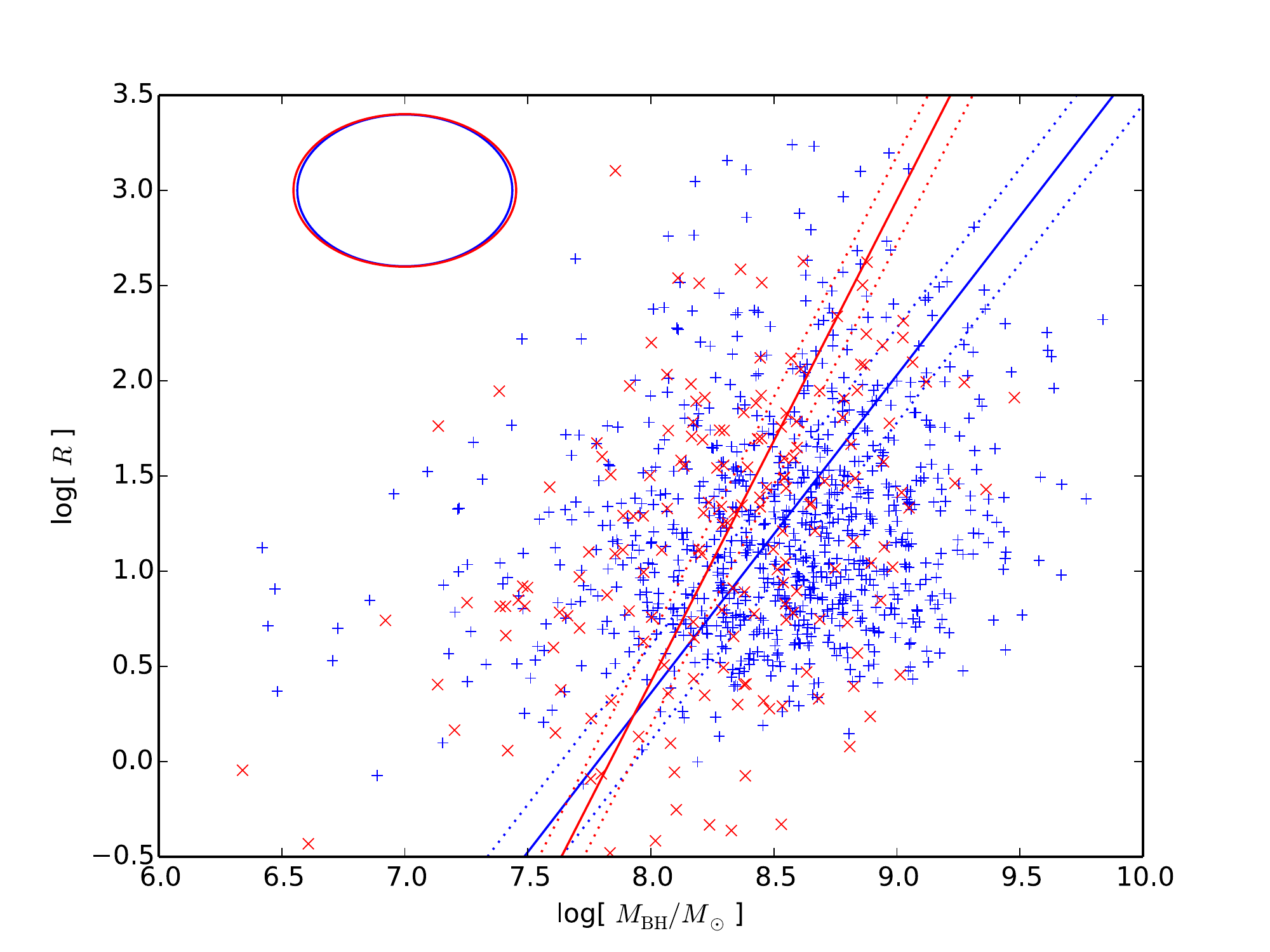} 
\caption{ \small
The same as Figure 1, but now showing the radio loudness $R$ versus the black hole mass for HERGs (red crosses) and LERGs (blue plusses). The best-fit relation is log $R$ = ($0.4\pm0.2$) + ($2.5\pm0.7$) [log $M_{BH} + 8]$ with $0.23\pm0.14$ dex intrinsic scatter for HERGs, and log $R$ = ($0.4\pm0.3$) + ($1.7\pm0.5$) [log $M_{BH} + 8]$ with $0.25\pm0.08$ dex intrinsic scatter for LERGs. }
\label{Fig_R_MBH}
\end{center}
\end{figure}

In Figures \ref{Fig_LR_MBH} and \ref{Fig_R_MBH}, we plot the 5 GHz radio luminosity ($L_{\mathrm{5GHz}}$) and the radio loudness parameter ($R$) as a function of the black hole mass ($M_{BH}$), respectively. 
The majority of the objects are radio-loud according to the standard radio-loud/radio-quiet division based on the radio loudness parameter ($\log R = 1$) and the alternative division based on the absolute radio luminosity \citep[e.g.][$L \sim 10^{23}$ W/Hz at 1.4 GHz corresponding to $L \sim 5 \times 10^{38} \mathrm{erg s^{-1}}$ at 5 GHz]{Best_et_2005}. 
The possible existence of a black hole mass threshold dividing radio-loud and radio-quiet AGNs has been suggested in the literature, with a somewhat arbitrary limit set around $\sim 10^8 M_{\odot}$ \citep{Laor_2000, McLure_Jarvis_2004, Chiaberge_Marconi_2011}. 
We observe that most of the radio galaxies in Figures \ref{Fig_LR_MBH} and \ref{Fig_R_MBH} have large black hole masses ($M_{BH} \gtrsim 10^8 M_{\odot}$). 
Although radio-loud objects seem to be typically associated with more massive black holes, the previously reported detection of a sharp boundary at $M_{BH} \sim 10^8 M_{\odot}$ is not evident in our sample. 
The fact that the majority of our sample is radio-loud and skewed towards large black hole masses is partly due to our sample definition based on radio selection. In fact, sample selections based on strong radio emission tend to yield the most massive black holes, and presumably the most massive host galaxies \citep{Best_et_2005}, a trend that has also been noted in \citet{McLure_Jarvis_2004}. 
As previously observed by \citet{B_H_2012}, HERGs typically tend to have lower black hole masses compared to the LERG population. 



\subsection{Dependence on the Eddington ratio}

\begin{figure}
\begin{center}
\includegraphics[angle=0,width=0.5\textwidth]{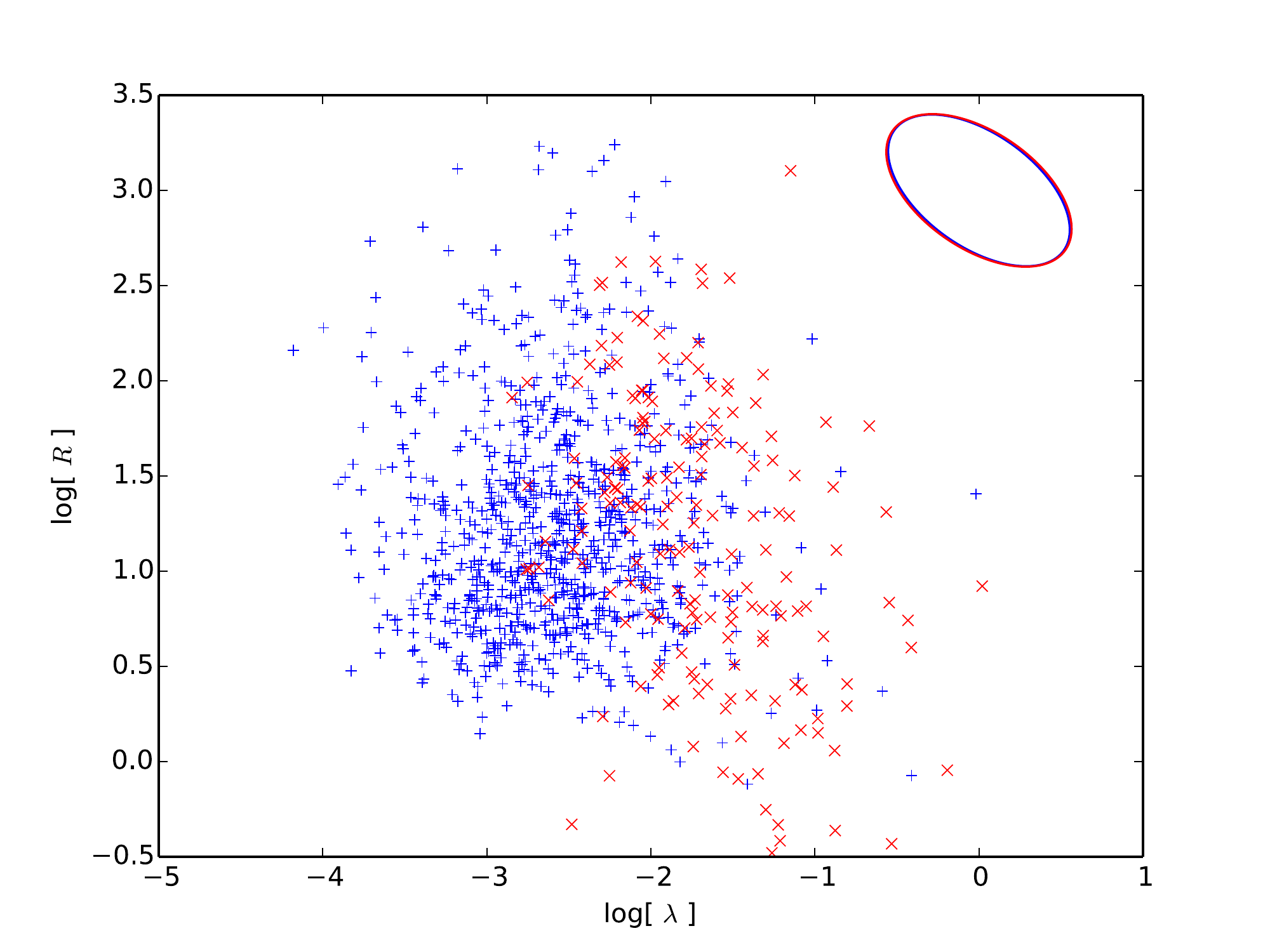} 
\caption{ \small
The radio loudness $R$ versus Eddington ratio defined as $\lambda = 10 L_B/L_E$, with HERGs indicated as red crosses and LERGs as blue plusses. We have not fitted a linear relation to these data because our derived $L_B$ is likely to be affected by significant systematics, as discussed in the text. Note that there is also significant covariance between $\lambda$ and $R$, as both quantities depend on $L_B$ and the statistical uncertainty on $L_B$ tends to be very large.
}
\label{Fig_R_lambdaB}
\end{center}
\end{figure}

\begin{figure}
\begin{center}
\includegraphics[angle=0,width=0.5\textwidth]{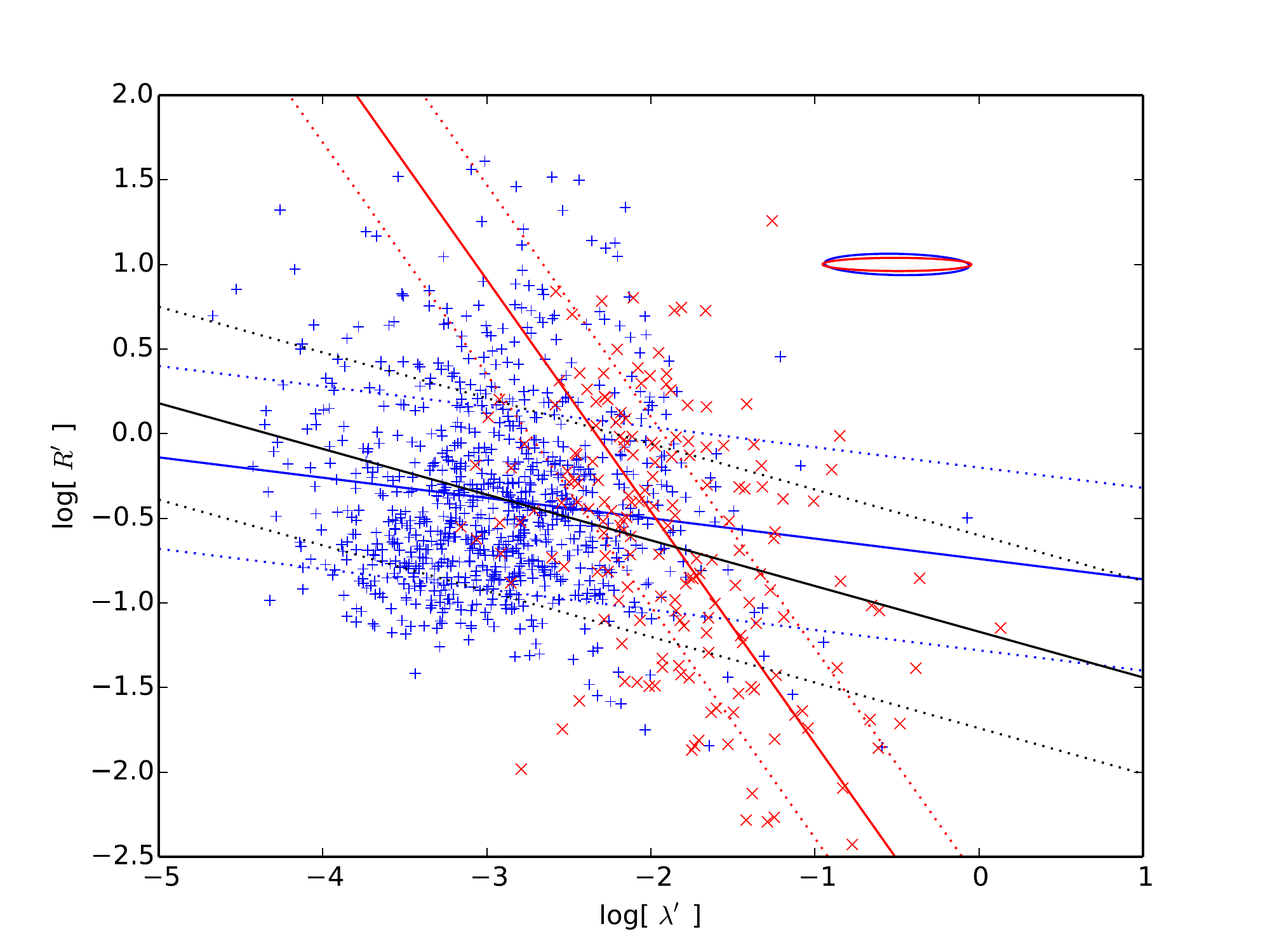} 
\caption{ \small
The same as Figure \ref{Fig_R_lambdaB}, but now showing the radio loudness defined as $R^{\prime}$ = L({\rm 1.4~GHz})/L({\rm [Oiii]}) versus the Eddington ratio defined as $\lambda^{\prime} = 3500 L({\rm [Oiii]})/L_{\rm Edd}$. 
The best-fit relation is log $R^{\prime}$ = ($-0.5\pm0.08$) + ($-1.4\pm0.4$) [log $\lambda^{\prime} + 2]$ with $0.56\pm0.08$ dex intrinsic scatter for HERGs, and log $R^{\prime}$ = ($-0.5\pm0.07$) + ($-0.1\pm0.07$) [log $\lambda^{\prime} + 2]$ with $0.54\pm0.01$ dex intrinsic scatter for LERGs. The black line gives the best-fit linear relation for the total population, log $R^{\prime}$ = ($-0.3\pm0.04$) + ($-0.6\pm0.04$) [log $\lambda^{\prime} + 2]$, with $0.57\pm0.01$ dex intrinsic scatter.}
\label{Fig_R_acc_orig}
\end{center}
\end{figure}

The radio-loudness parameter ($R$) as a function of the Eddington ratio ($\lambda$) is shown in Figure \ref{Fig_R_lambdaB}. 
A strong trend of increasing radio loudness with decreasing Eddington ratio has been reported in many previous works \citep[e.g.][]{Ho_2002, Sikora_et_2007, Sikora_et_2013}. 
As  already mentioned in Sect. \ref{Section_Data}, the way in which the B-band luminosities are derived here is not particularly reliable and introduces large uncertainties.  
It is therefore more relevant to define equivalent physical quantities in terms of parameters closer to direct observables, i.e. avoiding spurious conversions as far as possible. 
We thus introduce an alternative measure for the radio loudness parameter as the ratio of the 1.4 GHz radio luminosity to the [Oiii] line luminosity ($R' = L(\rm 1.4 GHz)/L(\rm [Oiii])$), and the Eddington ratio defined as $\lambda' = L_{\rm bol}/L_{\rm Edd}$, with the bolometric luminosity directly estimated from the [Oiii] line luminosity, $L_{\rm bol} = 3500 L(\rm [Oiii])$ \citep{Heckman_et_2004}.
In Figure \ref{Fig_R_acc_orig} we plot the radio loudness parameter ($R'$) as a function of the Eddington accretion rate ($\lambda'$), based on the new defined quantities. 
This plot is analogous to Figure \ref{Fig_R_lambdaB}, and in particular it is directly comparable to corresponding plots presented in \citet{Sikora_et_2013}. 
We observe that the radio-AGN sample is dominated by sources with low Eddington ratios, mainly $\log \lambda' < -2$, with a clear lack of sources approaching the Eddington limit ($\log \lambda' \sim 0$). 
The prevalence of low accretion rate objects already suggests that the local radio-AGNs are likely powered by some form of radiatively inefficient accretion flow. 
In fact, the majority of the radio galaxies in our sample are classified as LERGs. We also observe that LERGs typically have lower Eddington ratios compared to HERGs; this important difference in the Eddington-scaled accretion rate is one of the main drivers that has led to the picture of the dichotomy in the present-day AGN population \citep{B_H_2012}.

From Figure \ref{Fig_R_acc_orig}, we see that the density of points rises sharply below some critical value of the accretion rate, $\lambda' \sim 0.01$. 
A trend of increasing radio loudness with decreasing Eddington ratio, compatible with the $R - \lambda$ anticorrelation reported in previous studies, can also be recovered in Figure \ref{Fig_R_acc_orig}. 
But the lack of sources in the low Eddington ratio and low radio loudness regime could also be due to radio faintness related to flux limits, and the observed correlations may in part be driven by observational selection effects. 
We note that distinct slopes of the linear fits are obtained for the two populations, with HERGs having a steeper dependence on $\lambda'$ than LERGs.
Our best fit linear relation to the total population of local radio galaxies, taking into account all uncertainties, covariances, and intrinsic scatter, yields $R' \propto (\lambda')^{-0.3\pm0.04}$. 
This can be compared with the slope of $\sim -0.5$ found for a sample of galactic nuclei covering a wide range of nuclear luminosities \citep{Ho_2002}. 

\section{Comparison to previous work}
\label{Section_Comparison}

The connection between radio emission and black hole parameters, i.e. black hole mass and accretion rate, has been investigated in many previous works. 
The correlation observed between jet and accretion powers, as traced by radio and emission line/mid-infrared luminosities, suggests a close coupling hinting at some common physical mechanism \citep[e.g.][and references therein]{Fernandes_et_2011}. 
But somewhat contrasting results have been reported in the literature \citep{Xu_et_1999, Laor_2000, Ho_2002, McLure_Jarvis_2004, Sikora_et_2007, Chiaberge_Marconi_2011, Sikora_et_2013}.

A positive correlation between radio emission and black hole mass, whereby radio-loud objects are associated with massive black holes, has been reported by several authors \citep{Laor_2000, McLure_Jarvis_2004, Chiaberge_Marconi_2011}.
Based on a sample of 87 low-redshift ($z < 0.5$) Palomar-Green quasars, \citet{Laor_2000} claims a radio loudness bimodalilty directly linked to the black hole mass: most of the quasars with $M_{BH} > 10^9 M_{\odot}$ are radio-loud, whereas the majority of quasars with $M_{BH} < 3 \times 10^8 M_{\odot}$ are radio-quiet. 
The correlation between radio emission and black hole mass has been confirmed in a much larger sample of optically selected quasars from the SDSS \citep{McLure_Jarvis_2004}. A significant correlation between absolute radio luminosity and black hole mass has also been claimed for this sample.
The mean black hole mass of radio-loud quasars is found to be larger than that of the radio-quiet counterparts, with most of the radio-loud objects having $M_{BH} \geq 10^8 M_{\odot}$. 
Indeed, the fraction of radio-loud quasars with $M_{BH} \leq 10^8 M_{\odot}$ only forms a small minority of the luminous SDSS quasar population. The fraction of radio-loud objects with $M_{BH} \leq 10^8 M_{\odot}$ is also a minority in our local radio-AGN sample, although the nature of the underlying accretion flows is expected to be quite different in the two populations.
More recently, \citet{Chiaberge_Marconi_2011} investigate the relation between radio loudness and black hole mass, using up-to-date mass estimates, which explicitly take into account the effects of radiation pressure. 
From the analysis of three selected samples, consisting of low-luminosity AGNs, QSOs, and SDSS AGNs, they conclude that there are basically no radio-loud AGNs with $M_{BH} < 10^8 M_{\odot}$. 
The above studies suggest that a large black hole mass is a required element in order to produce a powerful radio source. 

On the other hand, several studies report a significant anti-correlation between radio loudness and Eddington ratio \citep[e.g.][]{Ho_2002, Sikora_et_2007, Sikora_et_2013}, compatible with the trends observed for our local radio-AGN sample. 
A strong anti-correlation of the radio loudness with the Eddington ratio, and hence accretion rate, has been found for a sample covering a large range of luminosities from inactive nuclei up to luminous quasars \citep{Ho_2002}.
The radio loudness is found to systematically increase with decreasing Eddington ratio, such that weakly accreting objects always tend to be radio-loud, while most of the high accretion rate objects are radio-quiet. 
This is also observed in the local radio-AGN sample, where most of the objects have low Eddington ratios and tend to be radio-loud. 
\citet{Sikora_et_2007} analyse a heterogeneous sample covering several orders of magnitude in both radio and optical luminosities, and observe that AGNs form two distinct sequences in the $L_{\mathrm{5GHz}} - L_B$ plane separated by three orders of magnitude in radio luminosity. The trend of increasing radio loudness with decreasing Eddington ratio is followed separately by both sequences, with the same dependence. 
More recently, \citet{Sikora_et_2013} report a strong anticorrelation between radio loudness and Eddington ratio for a sample of narrow-line radio galaxies at $z < 0.4$, confirming the negative trend found in earlier studies. 
Based on the observed systematic increase of the radio loudness with decreasing Eddington ratio, \citet{Sikora_et_2007} propose a revised radio-loud/radio-quiet division depending on the accretion rate: $\log R = - \log \lambda + 1$ for $\log \lambda > -3$ and $\log R = 4$ for $\log \lambda < -3$. 
According to this new variable boundary criterion, essentially all of our sources would fall in the radio-quiet category, suggesting that we are not actually sampling the most powerful radio sources. 

The local radio-AGN sample analysed here is selected by cross-matching the SDSS with the NVSS/FIRST radio catalogs, resulting in a larger size sample compared to many previous works. 
The division into high-excitation and low-excitation sources is generally based on emission line properties \citep[][and references therein]{B_H_2012}. 
For instance, an excitation index defined by a set of 6 emission lines has been used to split the 3CR radio galaxies sample into two distinct sub-populations \citep{Buttiglione_et_2010}. Although such divisions are based on optical spectroscopic classification, high-excitation and low-excitation galaxies are also found to be distinct in terms of their ratio of emission line to radio luminosities. 
In fact, for a given radio luminosity, HERGs are observed to be brighter in the [Oiii] emission line luminosity than LERGs by a factor of $\sim 10$, implying differences in their radio loudness characteristics.

Overall, the observed trends with black hole mass and accretion rate seem to be broadly consistent with results reported in previous studies.
However, we note that the observational samples discussed in this Section are selected according to different selection criteria (e.g. optically-selected vs. radio-selected) and are likely to be powered by different underlying accretion modes (see Discussion). 
For instance, the sample of optically selected quasars in the redshift range $0.1 < z < 2.1$, analysed by \citet{McLure_Jarvis_2004}, is mainly composed of radiatively efficient sources accreting at considerable fractions of the Eddington rate. 
On the other hand, most of the radio galaxies observed in the local Universe emit relatively little radiation, with much of the energy output released in kinetic form. Such sources are thus considered to be in the radiatively inefficient, jet-dominated regime \citep{B_H_2012}. 

In addition, one may also need to include a note of caution. 
While we try to carefully take into account measurement uncertainties and error propagation, there are intrinsic issues related to the way in which different physical parameters are plotted. 
In particular, one is concerned with the $R - \lambda$ representation, in which the same physical quantity appears on both axis as ratioed quantities. 
The resulting biases might affect previously reported results, although this aspect has not been emphasized much in the literature. 
Furthermore, one should consider the orientation of the source with respect to the observer, with the associated beaming effect, leading to related issues on the use of the core or total radio luminosities \citep[as discussed in e.g.][]{Broderick_Fender_2011}. 
With these caveats in mind, we attempt to provide possible physical interpretations of the observed trends in the next section.  


\section{Discussion}
\label{Section_Discussion}

We have analysed the dependence of the radio emission on black hole mass and Eddington ratio, based on a large sample of radio sources drawn from the SDSS. 
Most of the local radio-AGNs host massive black holes, typically in the mass range $M_{BH} \sim 10^8 -10^{9} M_{\odot}$. 
This suggests that radio galaxies are drawn from the population of massive galaxies, likely associated with giant elliptical galaxies in the local Universe. 
These giant ellipticals may be low-power radio sources, emitting relatively little radiation at optical wavelengths.
We note that our radio selection preferentially selects the most massive black holes with low accretion rates, i.e. radiatively inefficient AGNs in the radio mode \citep{B_H_2012}. 

In the $R$ vs. $M_{BH}$ plot, we note a lack of objects with small black hole mass and high radio loudness for the radio-selected sample. 
Although radio-loud AGNs are preferentially associated with massive black holes, for a given black hole mass, the range covered in both absolute radio luminosity and radio loudness can be several orders of magnitude. 
Moreover, one should take into account the large uncertainties in the black hole mass estimates, which tend to considerably broaden the observed distributions. 
It has been argued that a large black hole mass is a necessary condition in order to form a powerful radio source, although it is clearly not a sufficient one. 
In physical terms, it is not clear why the relative jet power of a source should have a strong dependence on the mass of the central black hole, and it is also not simple to imagine a physical process explicitly scaling with the black hole mass. 
One may consider downsizing effects, whereby in the present-day Universe low-mass black holes are in a radiatively efficient regime, accreting at rates close to the Eddington limit. 

On the other hand, the dependence of the radio loudness on the accretion rate can be interpreted in terms of transitions in the underlying accretion flow modes. 
It is well known that at high Eddington ratios, accretion proceeds via a geometrically thin, optically thick accretion disc, with a high radiative efficiency. This is the standard radiatively efficient accretion mode \citep{S_S_1973}. 
However, when the accretion rate drops below a certain critical value, the accretion flow can switch to a radiatively inefficient form (radiatively inefficient accretion flow or RIAF), such as the advection-dominated accretion flow (ADAF) \citep{Narayan_Yi_1994}. 
An ADAF is generally characterised by a geometrically thick and optically thin accretion flow.
Most of the energy released by viscous dissipation is retained in the gas and advected inwards, with the resulting radiative efficiency being very low.  
In this radiatively inefficient regime, the bulk of the energy is emitted in kinetic form through radio jets, and the spectral energy distribution of ADAFs is generally expected to be radio-loud with a lack of a strong blue bump component. 
The geometrical thickness of the inner torus, possibly surrounding a spinning black hole, may also help in the initial collimation of the radio jet \citep{Rees_et_1982, Fabian_Rees_1995}. 
Therefore the observed increase in the radio loudness with decreasing accretion rate can be understood in terms of two effects: an actual enhancement of the radio jet emission for low accretion rates and/or a decrease in the optical emission due to the decline of the radiative efficiency in the RIAF mode. 
Both effects may combine to produce the observed anticorrelation, as also noted by \citet{Wu_et_2013}. 

A similar pattern is actually observed in the case of black hole binaries, in which the presence (or absence) of a radio jet is closely coupled with the accretion state \citep{Remillard_McClintock_2006}. The low-hard state, operating at low accretion rates, is usually associated with a steady radio jet; while in the high-soft state, corresponding to high accretion rates, the radio jet seems to be quenched. 
In black hole binaries, the state transitions and associated levels of radio jet emission are thus interpreted in terms of a change in the underlying accretion modes.  
In this framework, the radio-loud AGNs in our local sample may be considered to reside in a state analogous to the low-hard state of black hole binaries.

The transition between the radiatively efficient accretion regime and the radiatively inefficient flow typically occurs around one per cent of the Eddington rate ($\lambda \sim 0.01$), below which the kinetic output dominates over the radiative output \citep[e.g.][]{Churazov_et_2005}. 
Most of the objects in our radio-AGN sample have low Eddington ratios (typically $\lambda < 0.01$), and are likely to be powered by radiatively inefficient accretion, with most of the energy released in kinetic form. 
In this context, the `radio loudness' of a source may be considered as a ratio of kinetic to radiative powers: 
\begin{equation}
R \propto \frac{L_{kin}}{L_{rad}} \propto \frac{\lambda_{kin}}{\lambda_{rad}}
\end{equation}
where $\lambda_{rad} = \epsilon_{rad} \dot M c^2 /L_E$ and $\epsilon_{rad}$ is the radiative efficiency, which actually depends on the nature of the underlying accretion flow.
In particular, below a certain critical accretion rate ($\dot{m}_{crit}$), the radiative efficiency scales as $\epsilon_{rad} \propto \dot{m}/\dot{m}_{crit}$, where $\dot{m} = \eta \dot{M} c^2 /L_E$ with $\eta$ the accretion efficiency \citep{Hopkins_et_2006, Merloni_Heinz_2008}. 
Assuming that the jet power scales linearly with the accretion rate ($\lambda_{kin} \propto \dot{m}$), and that the radiative output has a quadratic dependence on the accretion rate, of the form $\lambda_{rad} \propto \dot{m}^2$ \citep{Mocz_et_2013}, we obtain: $R \propto \dot{m}^{-1}$. 
In a naive picture, this may explain the trend of the kinetic and radiative outputs with the accretion rate. 
It is interesting to note that, based on the so-called `magnetically arrested' accretion flow scenario, \citet{Sikora_et_2013} predict a relation of the form $R \propto \lambda^{-0.4}$.  

However, the accretion rate cannot be the only parameter determining the relative importance of the jet power in AGNs. 
In fact, for a given Eddington ratio, we observe a huge difference in radio loudness, e.g. a factor of $\sim 10^3$ between the radio-loud and the radio-quiet sequences \citep{Sikora_et_2007}. 
Therefore a further physical parameter is required, in addition to mass and accretion rate, and black hole spin has been suggested as a potential candidate. We recall that radio emission is implicitly coupled to the relativistic jet component, which is in turn coupled to the black hole spin, hence a link between radio loudness and spin distributions has been searched for. 
This has led to the so-called `spin paradigm', according to which the presence of a powerful jet, and associated radio emission, is determined by the black hole spin \citep{B_Z_1977}. 
The other important parameter is the strength of the magnetic flux threading the central black hole. 
It has been argued that the required magnetic flux accumulation can only occur in geometrically thick accretion flows, and in such a configuration one can obtain powerful jets \citep[][and references therein]{Sikora_et_2013}. 
Therefore the association of a high black hole spin and a geometrically thick flow can produce the most powerful radio sources \citep[see also][]{Meier_2001}. 
Most likely, it is the combination of the three primary parameters, i.e. mass, accretion rate, and spin, that determines the radio loudness of a given source and its observational appearance. 
In addition, the underlying accretion flow geometry and the surrounding environment may also play an important role. 

As already mentioned, the relative importance of radiative versus kinetic feedback modes seems to be regulated by the accretion state. At high accretion rates, close to the Eddington limit, the radiative efficiency is high and the radiative output dominates over the kinetic output \citep{Churazov_et_2005}.
However, at low accretion rates the radiative efficiency drops dramatically, and the bulk of the energy is released in kinetic form via radio jets. 
The latter mode of feedback, characterised by low radiative efficiency, is thought to be the primary mode operating in the most massive galaxies in the present-day Universe. 
Indeed our sample of local radio-AGN is dominated by massive black holes accreting at very low fractions of the Eddington rate.
This supports a picture in which local radio galaxies, powered by radiatively inefficient accretion, are currently in the regime of radio mode feedback. 
We are clearly not sampling the most powerful radio sources, but rather the lower level radio AGN activity, which is more common and widespread in the local population. 
The transition between radiation-dominated and jet-dominated feedback modes has been discussed in terms of physical changes in the underlying accretion flows, from a radiatively efficient accretion disc to some form of radiatively inefficient flow. 
In an evolutionary scenario, a two-stage feedback process has been suggested, whereby the radiative mode dominates at high redshifts and builds the black hole-host galaxy relations, while at lower redshifts a stable state is maintained by the dominant kinetic mode \citep{Churazov_et_2005}. 
The next step will be to explicitly study the relation between radio mode feedback and the host galaxy properties, including stellar population parameters and environment. 


\section*{Acknowledgements}
WI acknowledges support from the Swiss National Science Foundation.   

\bibliographystyle{mn2e}
\bibliography{biblio.bib}


\label{lastpage}

\end{document}